\documentclass{article}
\usepackage{arxiv}
\usepackage{authblk}
\usepackage[utf8]{inputenc} 
\usepackage[T1]{fontenc}    
\usepackage{url}
\usepackage{booktabs}
\usepackage{threeparttable}
\usepackage{multirow}
\usepackage{subcaption}
\usepackage{amsfonts}
\usepackage{microtype}
\usepackage{graphicx}
\usepackage{siunitx}
\usepackage{physics}
\usepackage{interval}
\usepackage{doi}
\usepackage{nameref}
\usepackage[nameinlink,capitalise]{cleveref}
\usepackage[style=numeric-comp,sorting=none,isbn=false,url=false,doi=false]{biblatex}
\intervalconfig{soft open fences}

\DeclareSIUnit{\px}{\text{px}}

\newcommand{\eg}{\emph{e.g.} }
\newcommand{\ie}{\emph{i.e.} }

\title{A neural operator-based surrogate solver for free-form electromagnetic inverse design}
\author[1]{Yannick Augenstein}
\author[2]{Taavi Repän}
\author[1,3]{Carsten Rockstuhl}

\affil[1]{Institute of Theoretical Solid State Physics, Karlsruhe Institute of Technology, 76131 Karlsruhe, Germany}
\affil[2]{Institute of Physics, University of Tartu, Tartu 50411, Estonia}
\affil[3]{Institute of Nanotechnology, Karlsruhe Institute of Technology, 76021 Karlsruhe, Germany}

\begin{document}

\maketitle
\begin{abstract}
Neural operators have emerged as a powerful tool for solving partial differential equations in the context of scientific machine learning. Here, we implement and train a modified Fourier neural operator as a surrogate solver for electromagnetic scattering problems and compare its data efficiency to existing methods. We further demonstrate its application to the gradient-based nanophotonic inverse design of free-form, fully three-dimensional electromagnetic scatterers, an area that has so far eluded the application of deep learning techniques.
\end{abstract}

\section{Introduction}

Tools for solving Maxwell's equations are essential in nanophotonics for modeling light-matter interaction at the wavelength scale and, in extension, for designing new optical devices.
Some of the most established methods for this purpose are finite element solvers (FEM)~\cite{volakis1998finite,jin2015finite} in the frequency domain and finite difference solvers for both frequency (FDFD)~\cite{zhao2002compact,hughes2019forward} and time domain (FDTD)~\cite{oskooi2010meep,vaccari2011parallel,taflove2013advances}.
As full-wave Maxwell solvers, these methods represent the most general and accurate class of tools for modeling electromagnetic systems.
However, full-wave solutions often involve significant time and computational cost, placing a practical limit on the scale of problems that can be tackled.
This is exacerbated in problems such as inverse design~\cite{molesky2018inverse,hughes2018adjoint,schneider2019benchmarking,augenstein2020inverse,lin2020end}, where typically on the order of hundreds of simulations have to be performed to reach reasonable solutions.
While there are ongoing developments in the search for faster full-wave solvers~\cite{hughes2021perspective,lin2022fast}, there exist also a variety of semi-analytical methods~\cite{li1997new,liu2012s4,minkov2020inverse,beutel2021efficient} that can potentially offer order-of-magnitude speedups.
However, such methods make specific physical assumptions, and their applicability is, therefore, generally limited to certain problem classes.

A more recent development has been the use of machine learning-based surrogate models~\cite{pestourie2021physics,lu2021learning,chen2021improved} to approximate solutions to partial differential equations (PDEs).
Such models can -- during inference -- be as fast or faster than semi-analytical methods and can, in principle, be tailored to suit a wide variety of problems because of their property as universal approximators~\cite{hornik1989multilayer,lu2021learning}.
Machine learning has garnered significant momentum in nanophotonics in recent years~\cite{baxter2021deep,krasikov2022intelligent}, and surrogate solvers have been applied to both forward modeling and inverse design~\cite{an2019deep,jiang2019simulator,wiecha2019deep,pestourie2020active,chen2022high}.
Of course, such models bring their own challenges and limitations, among which the main ones are loss of accuracy and data efficiency.
As data efficiency implies using less data to achieve more, these two quantities have an inverse relationship -- at the cost of using more data, a model can generally be trained to produce higher accuracy predictions, and vice versa.
As high-quality training data is usually generated using classical methods, \eg full-wave solvers, it is crucial to consider this trade-off in any application that uses surrogate solvers since they quickly lose their edge over conventional methods when the cost of generating training data limits them.
This is exacerbated by the fact that surrogate models can not serve as a general-purpose tool for solving scattering problems -- instead, they are limited in scope to the type of problem modeled in the training data.
It is, therefore, essential to increase the data efficiency of such models and find applications where the use of surrogate solvers can maximally offset the cost of data generation.

The key to improving the data efficiency of surrogate solvers lies in the development and application of new models in the context of scientific machine learning.
One aspect that has proven effective is imparting some partial physical knowledge to the surrogate model~\cite{karniadakis2021physics}.
This can either be done by explicitly including governing equations into the model formulation, such as in the case of physics-informed neural networks (PINNs)~\cite{lu2021physics,cuomo2022scientific} or implicitly via penalizing unphysical solutions during training~\cite{lim2022maxwellnet,chen2022high}.
The latter is not tied to specific models and can generally be applied when the governing equations (or some of their properties) are known.
Parallel to the incorporation of physics information into machine learning models has been the development of new classes of model architectures.
In particular, learning operator mappings between function spaces using deep neural networks has been repeatedly shown to outperform previous approaches on PDE-constrained problem sets~\cite{li2020neural,li2020fourier,kovachki2021neural,lu2022comprehensive}.
Such models include graph kernel networks (GKN)~\cite{li2020neural}, deep operator networks (DeepONets)~\cite{lu2021learning,lu2022multifidelity}, and Fourier neural operators (FNO)~\cite{li2020fourier}, among others.

However, even the most data-efficient models can not offset the cost of data generation if the problem at hand requires only a handful of simulations in the first place.
Therefore, we identify inverse design~\cite{elsawy2021multiobjective,roques2022toward,wiecha2022inverse,yao2022trace} -- specifically gradient-based -- as a discipline that is well-suited to benefit from the speed of surrogate models and suffers little from their drawbacks~\cite{chen2022high, chen2022algorithm}.
In gradient-based inverse design, a functional element is optimized by incrementally maximizing some figure of merit.
Gradients of this figure of merit with respect to incremental changes in the geometry are then used to refine the device until an optimum is found iteratively.
Conventional inverse design, \eg topology optimization~\cite{sigmund2013topology}, typically takes a few hundred iterations to converge, where each iteration entails two full-wave simulations -- one for evaluating the figure of merit and one for obtaining its gradients via the adjoint method.
Depending on the problem, a single optimization in 3D can easily take days to complete, and the solutions generally depend strongly on the initial conditions, \ie ideally, many such optimizations should be performed to judge the quality of the optimization.
Further, such optimizations generally can not be parallelized, as the incremental updates are serial in nature.
As such, the only remaining possibility to accelerate the inverse design process is to speed up the individual simulations.

Neural network-based surrogate solvers seem ideally suited to tackle this challenge -- their inference time is negligible when compared to full-wave simulations, training samples are independent and can be generated in parallel, and they are inherently \emph{differentiable}, meaning they can generally be used as a drop-in replacement for differentiable Maxwell solvers~\cite{hughes2019forward,hammond2022high} in existing gradient-based optimization pipelines.
At the same time, their drawbacks are mitigated to an extent.
During an optimization, the simulation parameters are generally fixed, and only the device geometry varies, meaning that it is feasible to use specialized models trained on as little data as possible and stay within the training distribution during optimization.
Furthermore, for many inverse design problems, high numerical accuracy is often not crucial to reach a feasible solution, especially considering fabrication uncertainty that is typically on the order of a few percent.
In cases where high accuracy is required, an initial device design (obtained using low numerical accuracy) can often be refined using high-accuracy methods within a few additional iterations.

In this work, we implement a modified version of FNO as a surrogate solver for electromagnetic scattering problems.
The model is trained on a diverse set of free-form electromagnetic scatterers. We compare its performance to a state-of-the-art convolutional architecture (UNet) in a simplified two-dimensional setting. We show that FNO requires significantly less data to reach the same accuracy as UNet.
In the~\nameref{inverse}, we use this model for the gradient-based inverse design of free-form scatterers in three dimensions from many different initial conditions, a task that is extremely computationally demanding using full-wave solvers.

We provide full access to our code as well as training datasets and model weights in accordance with the FAIR data principles~\cite{wilkinson2016fair}, see the~\nameref{availability} for further details.

\section{Fourier neural operator for electromagnetic field inference}

\begin{figure}[h]
    \centering
    \includegraphics{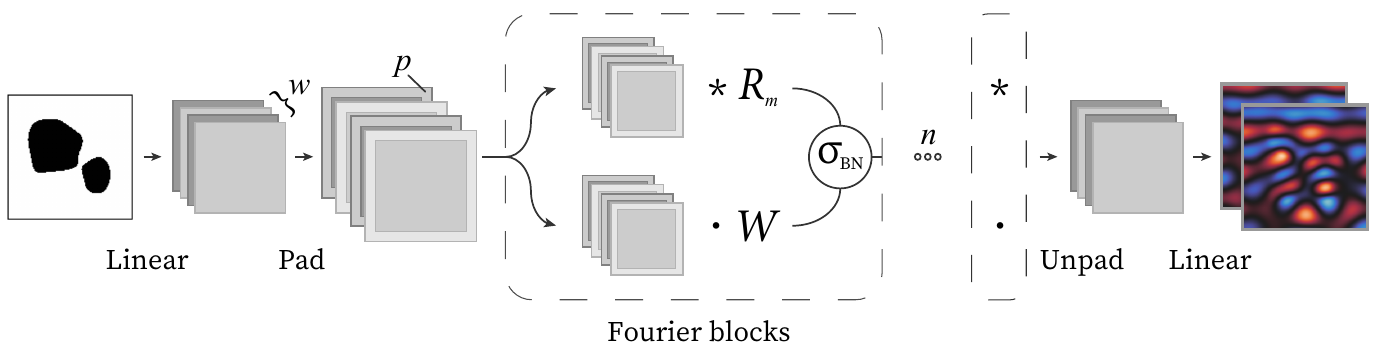}
    \caption{Illustration of the FNO architecture. The input is expanded via a linear layer to $w$ channels and then zero-padded by $p$ pixels. Data is then passed through a series of $n$ ``Fourier blocks'', each containing a linear layer and a convolution in Fourier space using the learned kernel $R_m$, a batch normalization layer as well as a GELU nonlinearity. Finally, the padding is removed, and the channel dimensions are reduced via a linear layer to the desired output dimensions.}\label{fno_arch}
\end{figure}

We begin with a brief description of the vanilla Fourier neural operator (FNO) introduced in~\textcite{li2020fourier} and outline some architectural choices particular to this work.
The core idea behind the FNO architecture is learning a kernel parametrized in Fourier space, where each Fourier layer in the FNO performs a \emph{global} convolution on its input.
First, the input $v(x)$, where $x$ denotes a location on the computational mesh, is lifted to a higher dimensional representation $y(x)$ with
\begin{equation}
    y(x) = C_{\text{in}}\qty(v\qty(x)) \in \mathbb{R}^w \qq{,}
\end{equation}
where $C_{\text{in}}: \mathbb{R}^{d_\text{in}} \rightarrow \mathbb{R}^w$ is a transformation parametrized by a linear layer with the input dimension $d_{\text{in}}$ and the output dimension $w$.
While the dimension of the input $d$ is determined by the problem and is typically small, the output dimension $w$ is a hyperparameter termed ``width'' of the FNO and corresponds to the number of kernels (feature channels) in each Fourier layer.
After lifting the input dimension, the data is passed through a sequence of $n$ ``Fourier blocks''.
Each such block consists of the Fourier layer $\kappa_m(y)$, a linear update, and a batch normalization layer~\cite{ioffe2015batch} followed by an activation function (ReLU) $\sigma_{\text{BN}}$:
\begin{equation}\label{fourier_layer}
    u(y) = \sigma_{\text{BN}} \biggl(\underbrace{\mathcal{F}^{-1}\qty(R_m \cdot \mathcal{F}(y))}_{\kappa_m(y)} + W y + b\biggr) \qq{,}
\end{equation}
with the Fourier transform $\mathcal{F}$, its inverse $\mathcal{F}^{-1}$, and the trainable parameters $R_m$, $W$, and $b$.
Together, $W$ and $b$ form the linear layer update, where $W$ is a weight matrix that acts locally on $y$, and $b$ is the bias vector.
The complex-valued tensor $R_m$ represents the kernel matrix of the convolution in Fourier space.
After the Fourier blocks, $y$ is mapped to the desired output dimensionality $d_{\text{out}}$ via a linear layer:
\begin{equation}
    z(x) = C_{\text{out}}\qty(y\qty(x)) \in \mathbb{R}^{d_{\text{out}}} \qq{.}
\end{equation}

A defining feature of FNOs is that the Fourier layer $\kappa_m(y)$ truncates the Fourier series at the $m$-th Fourier coefficient, \ie $R_m$ only contains entries up to the frequency index of that mode.
This truncation has some important implications on the expressiveness of an FNO network.
Most importantly, it acts as a low-pass filter by construction, and high-frequency spatial components are strongly suppressed in the output, leading to relatively smooth outputs (depending on the choice of $m$).
It turns out that this is advantageous in the context of many PDEs that describe physical systems, in particular for systems that have wave-like solutions such as Maxwell's equations.
This makes FNOs uniquely suited to model these systems, as this property is an inherent part of the architecture and does not need to be enforced through other means, such as explicitly smoothing the output or modifying the loss function -- both approaches are technically feasible but introduce additional complexity.
Moreover, approaches that modify the loss function, such as the ones used in~\textcite{chen2022high} or~\textcite{lim2022maxwellnet}, which aim to make the model predictions more self-consistent with Maxwell's equations, can be readily incorporated in the training of FNOs if necessary.
Here, we are only concerned with establishing a baseline performance for FNOs in the context of electromagnetic scattering problems.

In practice, vanilla FNOs use \emph{feature expansion} on the inputs to achieve good accuracy~\cite{lu2022comprehensive,li2020fourier,kovachki2021neural}.
The canonical choice is to add a Cartesian coordinate grid as an additional input feature, \ie the coordinates in each spatial dimension are added to the input channels, and the input dimensionality becomes, \eg, $\mathbb{R}^d \rightarrow \mathbb{R}^{d + 3}$ in three dimensions.
We have found that similar accuracy can be achieved by zero-padding the input ($p = 2$) before passing it through the Fourier blocks and removing the padding again before the final layer of the network, eliminating the need for feature expansion.
Further, we use the Gaussian Linear Error Unit (GELU) activation function~\cite{hendrycks2016gaussian} instead of ReLU, which we have empirically found to lead to slightly lower prediction errors.
A pictorial representation of the FNO architecture used here is shown in~\cref{fno_arch}.

\section{Results}

Our aim is to use FNOs as surrogate solvers for Maxwell's equations in place of full-wave methods such as the finite-difference time-domain (FDTD) method for some particular scattering problem.
To do this, we define the input of the FNO to be a dielectric material distribution that is discretized on a regular grid, meaning that the number of input features $d_{\text{in}} = 1$.
The output should then contain the complex electromagnetic field components of interest, where each field component uses two output channels for the real and imaginary parts, respectively.
The FNOs discussed in this work are fully described by the set of hyperparameters in~\cref{fno_hparams}.
\begin{table}[h]
\centering
\caption{Hyperparameters describing the FNO architecture. The column ``value'' contains the parameter values for the networks discussed in this work (both in 2D and 3D).}\label{fno_hparams}
\begin{tabular}{llr}
\toprule
Parameter & Description & Value \\
\midrule
$n$ & No. of Fourier blocks & 10 \\
$m$ & Fourier modes (truncation order) & 12 \\
$w$ & Hidden channels (width) & 32 \\
$p$ & Zero-padding in each spatial direction & 2 \\
\bottomrule
\end{tabular}
\end{table}

To train the models, we generate datasets of scatterers and fields using a full-wave Maxwell solver.
A detailed description of the data generation is given in the~\nameref{data}.
As a training and evaluation metric, we use the normalized $L_p$ loss
\begin{equation}\label{loss}
    L_p(\hat{\vb*{y}}, \vb*{y}) = \frac{1}{n} \sum_{i=1}^n \frac{\norm{y_i - \hat{y}_i}_p}{\norm{y_i}_p}
\end{equation}
throughout.
The models are trained on the normalized $L_2$ loss, \ie normalized root mean square error.
However, we will generally refer to the normalized $L_1$ loss for evaluation and discussion due to its more intuitive interpretability as the absolute error between two samples.
Further, we note that the discrepancy between $L_1$ and $L_2$ is minimal for the examples in this work, and in practice, the two can be used almost interchangeably.
For evaluation, we use a separate test set containing 400 samples not included in the training data (neither for training nor validation).
When referring to specific models, we use the name ``FNO-2D'' for models trained on two-dimensional data and ``FNO-3D'' for models trained on three-dimensional data.
For remarks and observations that apply more generally, we will use the term ``FNO'' as before.

\subsection[field inference section]{Field inference}\label{forward}

Previous works have focused on UNet-like~\cite{ronneberger2015u} convolutional architectures for electromagnetic field inference~\cite{wiecha2019deep,chen2022high,lim2022maxwellnet}.
To make an objective comparison between UNet and FNO performance for this task, we use the UNet architecture from~\textcite{chen2022high} as a reference implementation and train both models on datasets of simulations of a diverse set of random scatterers in two dimensions and vary the number of training samples to judge their respective data efficiency, \ie the number of training samples needed to reach a certain prediction accuracy.
Here, both networks use two output channels that contain the real and imaginary parts of the $z$-component $E_z$ of the electric field, respectively.
Both models are trained under the same conditions, and details of the training procedure are given in the~\nameref{fno_unet_training}.

\begin{figure}[h]
    \centering
    \includegraphics{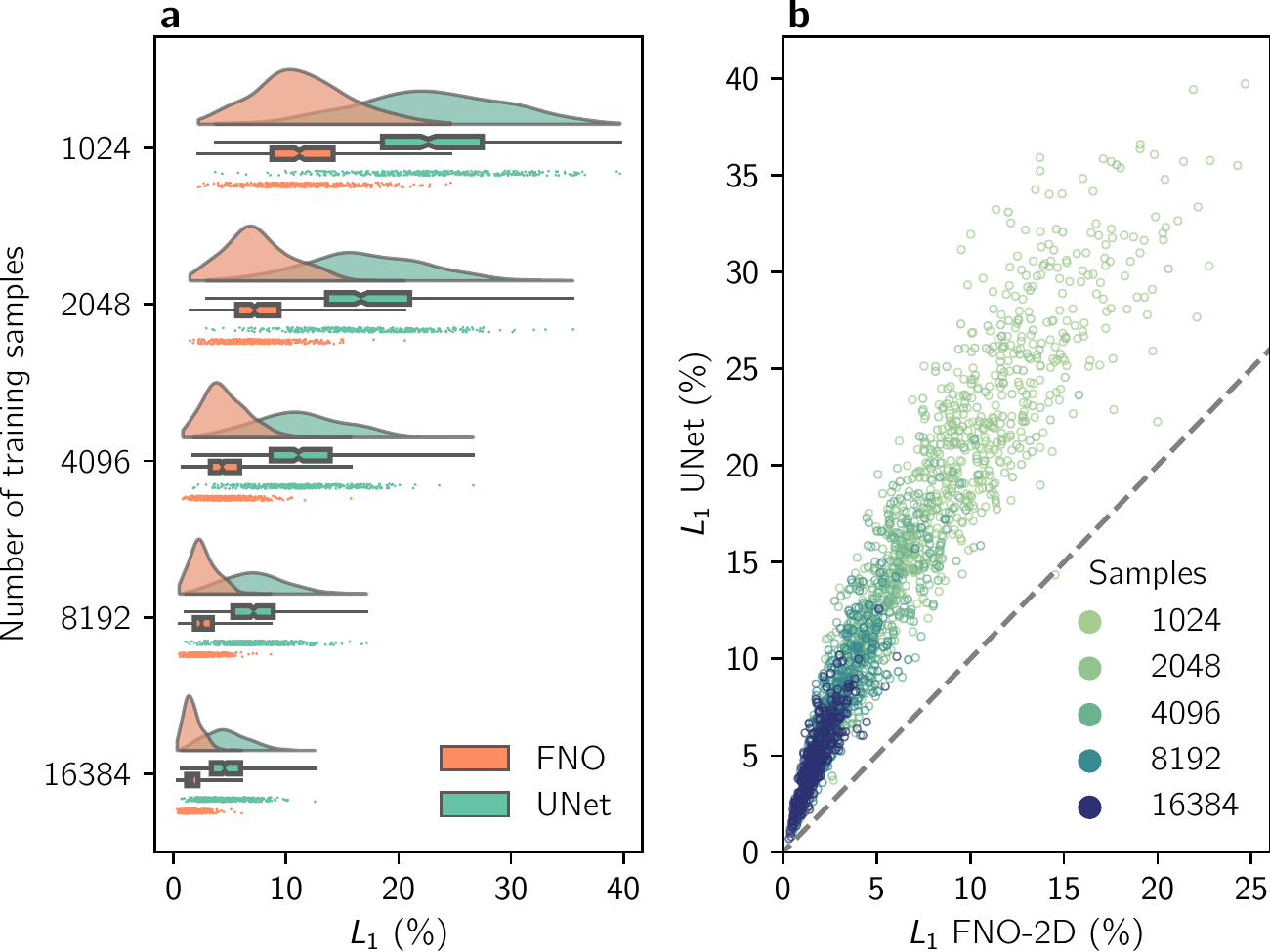}
    {\phantomsubcaption{\label{nets_raincloud}}}
    {\phantomsubcaption{\label{nets_scatter}}}
    \caption{\textbf{a} Raincloud plot~\cite{allen2019raincloud} of the normalized $L_1$ errors on a test set with 400 samples for FNO-2D and UNet trained on varying training set sizes. Each set of training samples shows a kernel density estimate (top, using Scott's rule for bandwidth estimation~\cite{scott1979optimal}), a box plot of the distribution with a notch indicating the median (middle), as well as a categorical scatter plot containing all samples (bottom). \textbf{b} Comparative scatter plot of the $L_1$ errors for all test samples and training set sizes of FNO-2D and UNet. The dashed gray line indicates where UNet and FNO-2D errors are equal.}\label{nets_dists}
\end{figure}

The evaluation results on the 400-sample test set are visualized in~\cref{nets_dists}, and numerical results are provided in~\cref{nets_results}.

\begin{table}[h]
\centering
\begin{threeparttable}
\caption{Median normalized $L_1$ and $L_2$ errors as measured over 400 test samples for networks trained on a varying number of training samples. The lowest error achieved on each training set size is marked \textbf{bold}.}\label{nets_results}
\begin{tabular}{rlrrrr}
\toprule
Samples & Network & $L_1$ (\%) & $L_2$ (\%) & $\sigma(L_2)$ (\%) & Time per epoch (s)\tnote{\textasteriskcentered} \\
\midrule
\multirow{2}{*}{$1024$}
& FNO-2D & $\mathbf{11.18}$ & $\mathbf{12.07}$ & $4.39$ & $8.31$ \\
& UNet & $22.64$ & $24.42$ & $6.80$ & $10.15$ \\ \rule{0pt}{2.3ex}
\multirow{2}{*}{$2048$}
& FNO-2D & $\mathbf{7.19}$ & $\mathbf{7.72}$ & $3.15$ & $10.10$ \\
& UNet & $16.68$ & $18.01$ & $5.59$ & $13.24$ \\ \rule{0pt}{2.3ex}
\multirow{2}{*}{$4096$}
& FNO-2D & $\mathbf{4.32}$ & $\mathbf{4.64}$ & $2.12$ & $13.72$ \\
& UNet & $11.09$ & $11.92$ & $4.15$ & $18.96$ \\ \rule{0pt}{2.3ex}
\multirow{2}{*}{$8192$}
& FNO-2D & $\mathbf{2.48}$ & $\mathbf{2.65}$ & $1.30$ & $21.13$ \\
& UNet & $7.10$ & $7.58$ & $2.88$ & $30.92$ \\ \rule{0pt}{2.3ex}
\multirow{2}{*}{$16384$}
& FNO-2D & $\mathbf{1.59}$ & $\mathbf{1.68}$ & $0.89$ & $35.81$ \\
& UNet & $4.52$ & $4.84$ & $1.99$ & $55.26$ \\
\bottomrule
\end{tabular}
\begin{tablenotes}
\item[\textasteriskcentered] Timing was performed on a single NVIDIA A100 SXM4 GPU.
\end{tablenotes}
\end{threeparttable}
\end{table}

While both network architectures see a significant reduction in test error for an increasing number of training samples, we observe that FNO-2D consistently outperforms UNet.
Not only does FNO-2D have significantly better prediction accuracy, but the distribution of errors on the test set shows a much tighter spread around the mean, indicating that FNO-2D not only has a higher prediction accuracy overall but also generalizes better to samples that are far from the dataset mean.
This is particularly remarkable considering that FNO-2D has only \num{5909250} trainable parameters in total, compared to UNet, which has \num{23617970} trainable parameters, meaning that FNO-2D is significantly more parameter-efficient.
To reach a test error of around \SI{5}{\percent}, FNO-2D requires 4096 training samples, whereas UNet requires four times as many to reach a comparable value.
As shown in~\cref{nets_raincloud}, this observation holds over the whole range of training set sizes that we considered, where UNet generally lags behind FNO-2D performance by a factor of four with respect to training set size.
We note that these comparative results are broadly in line with previous findings on benchmarking FNO and UNet performance on PDE-constrained problem sets~\cite{li2020fourier,kovachki2021neural}.
Furthermore, for the same number of samples, FNO-2D training is roughly \SI{30}{\percent} faster than UNet, with the gap widening slightly for a larger number of training samples.
This is, however, mostly because FNO-2D has significantly fewer parameters than UNet -- in fact, it is slower than UNet when training time is considered per parameter.

It is worth investigating whether the lower test error of FNO-2D holds for all samples or if, while having a lower error on average, some samples are nonetheless better predicted by UNet.
Such a scenario might indicate that for specific geometries, choosing UNet over FNO-2D could still be advantageous.
A comparison between FNO-2D and UNet $L_1$ losses for each test sample and all training runs is visualized in the scatter plot in~\cref{nets_scatter}, where the dashed line marks the line where FNO-2D and UNet errors are equal.
From this, we can clearly see that FNO-2D outperforms UNet for \emph{all} test samples irrespective of training set size.

\begin{figure}[h]
    \centering
    \includegraphics{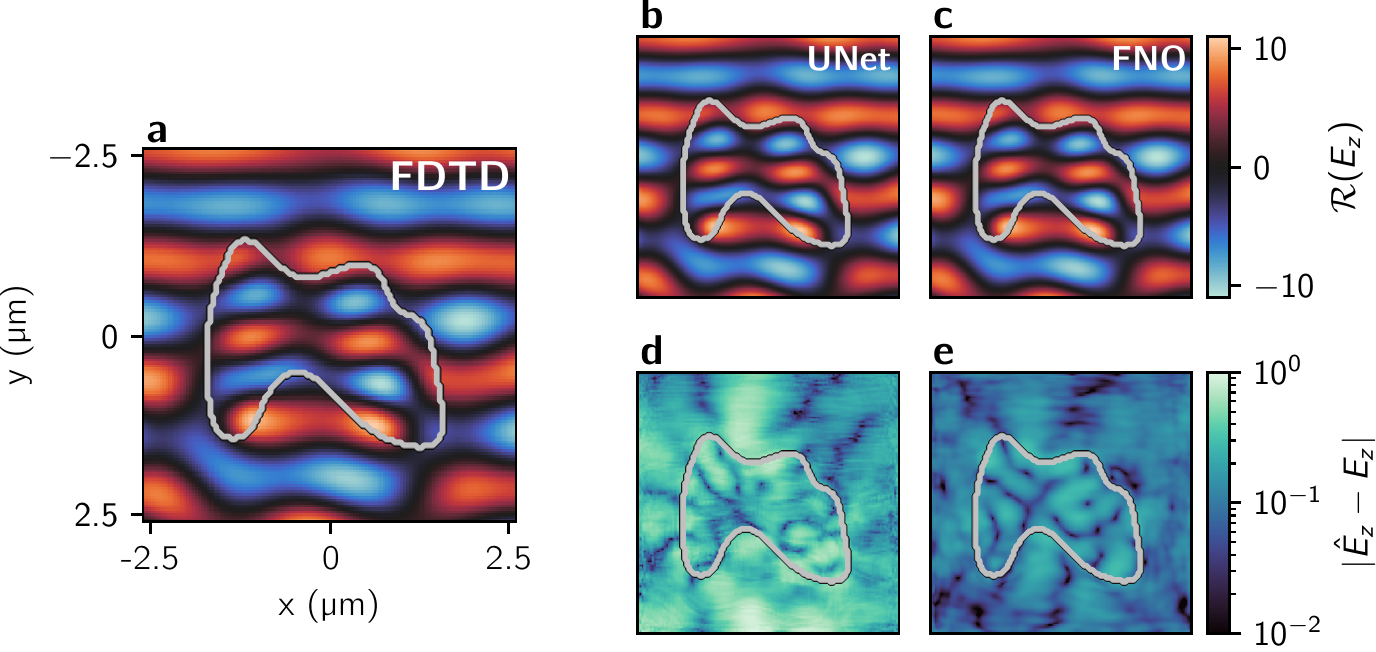}
    {\phantomsubcaption{\label{field_fdtd}}}
    {\phantomsubcaption{\label{field_unet}}}
    {\phantomsubcaption{\label{field_fno}}}
    {\phantomsubcaption{\label{err_unet}}}
    {\phantomsubcaption{\label{err_fno}}}
    \caption{Real part of the \textbf{a} ground truth, \textbf{b} UNet predicted, and \textbf{c} FNO-2D predicted electric field $E_z$ for a random scatterer drawn from the test set. Also shown are the absolute errors between the predicted fields and the ground truth for the \textbf{d} UNet prediction and \textbf{e} FNO-2D prediction. FNO-2D has a normalized $L_1$ error of \SI{1.58}{\percent}, and the UNet model has a normalized $L_1$ of \SI{4.29}{\percent} on the sample. Both networks were trained on the same dataset containing 16k training samples. The outline of the scatterer is shown in all plots. The samples are illuminated with a plane-wave source impinging from the negative y-direction.}\label{fields}
\end{figure}

\Cref{fields} shows a typical sample drawn from the test set, evaluated on the FNO-2D and UNet models trained on 16k samples.
The sample is chosen to be as close as possible to the median FNO-2D $L_1$ error on the test set and has a relative $L_1$ error of \SI{1.58}{\percent} for the FNO-2D prediction and an error of \SI{4.29}{\percent} for the UNet prediction.
We can see a good qualitative agreement between the fields in~\cref{field_fdtd,field_fno,field_unet} for both models.
This picture changes, however, when we consider the absolute error maps between the ground truth and the field predictions in~\cref{err_fno,err_unet}.
We observe that the UNet prediction has a significantly higher deviation throughout, which of course, is already clear from the data presented.
Somewhat more interestingly, we note that the error map produced by the FNO-2D model in~\cref{err_fno} shows less spatial noise when compared to the one produced by the UNet model in~\cref{err_unet}, \ie the map appears smoother overall.
This is in line with what one would expect from these two architectures -- while convolutional neural networks perform pixel-wise local convolutions, FNO instead performs \emph{global} convolutions and explicitly discards Fourier components above a certain order by design.
Essentially, the priors built into the FNO architecture force the model to learn continuous solutions, a property that is later reflected in its output.

\subsection[inverse design section]{Inverse design}\label{inverse}

We will now demonstrate the inverse design of free-form, three-dimensional nanophotonic devices using an FNO model trained on a dataset of $8192$ pairs of volumetric scatterers and fields.
The data generation and FNO-3D training are detailed in the~\nameref{data} and the~\nameref{fno_unet_training}.
After 100 epochs of training, the normalized $L_1$ loss of the FNO-3D model on the 400-sample test set is \SI{5.05}{\percent}.

\begin{figure}[h]
    \centering
    \includegraphics{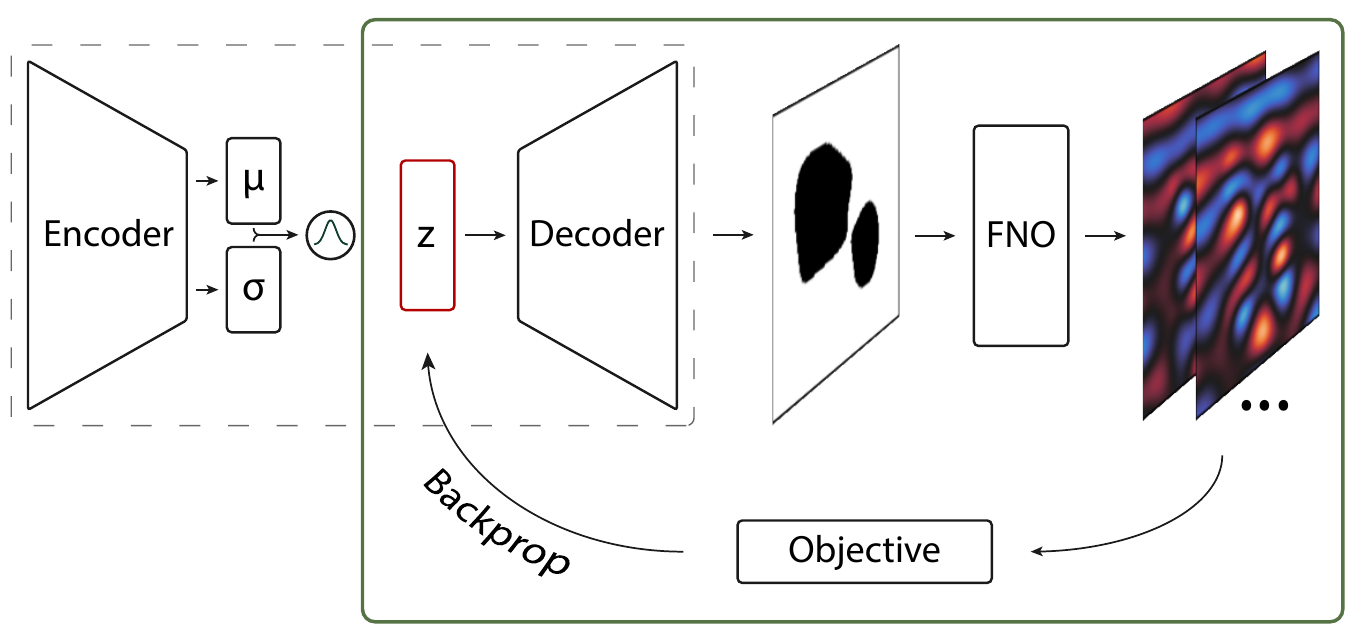}
    \caption{Flowchart of the iterative inverse design procedure. The decoder of a pre-trained variational autoencoder (dashed gray box) is used to generate scatterers from the latent vector $z$ (red box). The electric field is then obtained via the FNO model, from which the loss is calculated. Gradients of the objective with respect to the latent variables are obtained via backpropagation, from which they are then iteratively updated until the optimization converges.}\label{inverse_design}
\end{figure}

The inverse design process is relatively straightforward, as we use FNO-3D in place of a Maxwell solver -- the objective function can still be an arbitrary (differentiable) function of the fields.
However, our formulation differs from the typical density-based approach to inverse design in how the scatterers are parametrized.
Instead of optimizing the scatterers directly on the computational grid, they are represented in the latent space of a pre-trained variational autoencoder (VAE)~\cite{kingma2013auto}.
The reason for this is the gradient-based nature of the design process -- parametrizing the scatterers directly on the grid would lead to continuous variations in the permittivity of the scatterer.
However, we train FNO-3D only on binary data, meaning that field inference will fail for scatterers with permittivities outside of this range.
While it is straightforward to include more permittivity values in the training data, this would require a fine sampling between the lowest and highest permissible permittivity for continuous inverse design, which would lead to a significant increase in dataset size, defeating the purpose of using a specialized surrogate solver for \emph{fast} inverse design.
Instead, we train a convolutional VAE that generates random scatterers from the same distribution from which the training set was generated.
Afterward, the device of interest can be optimized directly in the latent space of the VAE.
Crucially, the latent space vector can be continuously updated, and the VAE decoder will map this to (approximately) binary geometries for which FNO-3D can give accurate predictions.
The encoder part of the VAE is only needed during training and is not used during inverse design.
Note that data generation for VAE training is very cheap as there is no simulation involved -- in fact, the data is generated on-the-fly during training.
Details of the VAE architecture and training are given in the~\nameref{vae_training}.
Note that the VAE parametrization is not a requirement for using FNO for inverse design -- in principle, any parametrization that yields binary scatterers can be used, \eg an explicit geometrical or boundary parametrization.
A pictorial representation of the inverse design pipeline is shown in~\cref{inverse_design}.

\begin{figure}[h]
    \centering
    \includegraphics{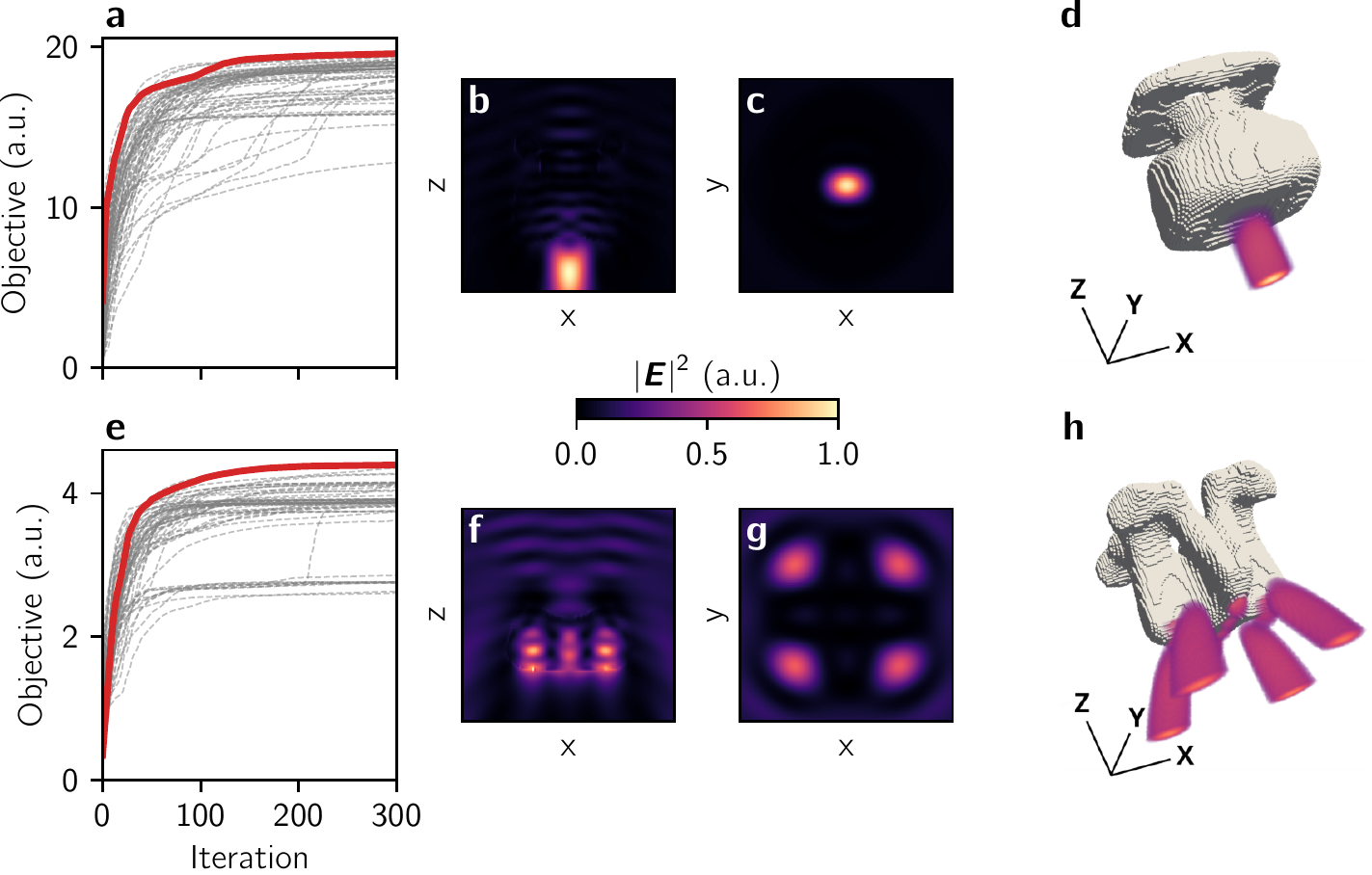}
    {\phantomsubcaption{\label{hist_1p}}}
    {\phantomsubcaption{\label{xz_1p}}}
    {\phantomsubcaption{\label{xy_1p}}}
    {\phantomsubcaption{\label{3d_1p}}}
    {\phantomsubcaption{\label{hist_4p}}}
    {\phantomsubcaption{\label{xz_4p}}}
    {\phantomsubcaption{\label{xy_4p}}}
    {\phantomsubcaption{\label{3d_4p}}}
    \caption{Inverse-design using FNO-3D of (\textbf{a}-\textbf{d}) a nanophotonic device that focuses into a single spot and (\textbf{e}-\textbf{h}) a nanophotonic device that focuses into four focal spots under plane-wave illumination. We show the (\textbf{a}, \textbf{e}) respective optimization histories of 64 optimizations for each device with different random initial conditions. The red line marks the best final device. Slices of the electric field intensity $|\vb*{E}|^2$ for the best devices as obtained from full-wave FDTD simulations are shown in the (\textbf{b}, \textbf{f}) $x$-$z$ plane as well as in the (\textbf{c}, \textbf{g}) focal $x$-$y$ plane. Images \textbf{d} and \textbf{h} show three-dimensional views of both devices along with a volume rendering of the field intensity thresholded such that only the highest-intensity regions are visible.}\label{inverse_design_results}
\end{figure}

As a demonstration, we optimize two nanophotonic devices using FNO-3D.
We define a simple objective function in terms of the electric field:
\begin{equation}
    J(\vb*{E}) = \sum_{\mathcal{D}} \qty|\vb*{E}(\vb*{r})|^2 \quad \forall \vb*{r} \in \mathcal{D} \qq{,}
\end{equation}
where the domain $\mathcal{D}$ represents the spatial points at which we wish to maximize the electric field intensity.
For the first device, we maximize the intensity at a single focal spot centered in the $x$-$y$ plane, \ie we design a simple nanophotonic lens.
In the second example, we maximize the intensity at four focal spots, each centered in one quadrant in the $x$-$y$ plane.
The plane wave illumination spans the $x$-$y$ plane and impinges from the top ($z = \SI{0}{\um}$) and the focal plane lies at the opposite end at $z = \SI{4.8}{\um}$.
We optimize both devices for 300 iterations using AdamW~\cite{loshchilov2017decoupled}.
In principle, other optimization algorithms such as L-BFGS-B~\cite{zhu1997algorithm} or MMA~\cite{svanberg1987method} can be used -- however, we choose AdamW here because of readily available GPU implementations and support for optimizing multiple devices in parallel.
The results for these optimizations are shown in~\cref{inverse_design_results}.

Crucially, we run 64 trials using different initial conditions -- random vectors drawn from a normal distribution in the VAE latent space -- to arrive at 64 different optimized devices for each problem, from which we then choose the best performing one as the ``champion device'' (see~\cref{hist_1p,hist_4p}).
We see that many runs converge to values significantly worse than the best one, \ie the optimization is highly sensitive to the choice of initial parameters, which is a problem that any local optimization faces.
However, sampling many different initial configurations is often not feasible for an inverse design using full-wave solvers, and only a few, if any, additional trials are typically performed.
In contrast, a single FNO-3D optimization takes around 10 minutes ($\approx 2$ seconds per iteration, NVIDIA A100 SXM4 GPU), and two optimizations can be run in parallel on a single GPU.
We run the optimizations for all 64 trials in parallel on 32 GPUs so that each full optimization run still takes only 10 minutes.
This, of course, depends on the computational facilities at hand, but even running all optimizations serially would take just over 5 hours.
For comparison, a single such optimization would take several days (20 minutes per simulation, two simulations per iteration, see the~\nameref{data}), rendering a sweep over many different initial conditions such as the one demonstrated here practically infeasible.

We perform full-wave simulations to evaluate the performance of the final optimized devices.
For the lens shown in~\cref{3d_1p}, the $L_1$ error between simulated and predicted field intensity $\qty|\vb*{E}|^2$ is \SI{10.2}{\percent}, and for the 4-point lens in~\cref{3d_4p}, the same error is \SI{9.1}{\percent}.
While this is slightly higher than the error on the test set, it is still well within reasonable accuracy, in particular regarding the uncertainty of microfabrication technologies, \eg 3D laser nanoprinting, that would be used to realize devices like this.
 The reason for the increased error when compared to the test error is simple - the error on the squared absolute fields is necessarily larger.
 In fact, the $L_1$ losses on $\qty|\vb*{E}|$ are in line with those of the test set means -- \SI{5.8}{\percent} for the single lens and \SI{5.0}{\percent} for the 4-point lens.
 Additionally, we note that the error in the intensity comes mainly from the absolute numerical values of the field components -- the qualitative agreement between FDTD and FNO-3D fields remains high, and we can see that both devices fulfill their respective design goals.
 All fields shown in~\cref{inverse_design_results} are taken from full-wave FDTD simulations.

\section{Discussion}

An (on the surface) compelling advantage of using surrogate models for solving scattering problems is the almost negligible inference time required when compared to running a full-wave solver.
However, there are some major concessions that lead to this speedup, and it requires careful consideration whether this trade-off is acceptable for a given task or not.
The most obvious caveats are perhaps generalization and accuracy -- outside of very specific cases~\cite{power2022grokking}, a surrogate solver trained on finite data will not generalize to data that lies outside of the distribution that it was trained on, and the inference error will only approach zero given sufficient network expressivity~\cite{raghu2017expressive} as well as training set size~\cite{goodfellow2016deep}.
More concretely, this means that for optical simulations, a data-driven model can only ``solve'' certain classes of scattering problems -- namely those that it was trained on, \eg specific source distributions, materials, and domain sizes, to name a few.
These limitations can generally be overcome to a degree by choosing a suitable network architecture, increasing the variety of training samples, and increasing the total amount of training data, leading to better generalization and increased accuracy.
However, if the cost of generating sufficient training data exceeds the cost of solving the task using conventional methods, then there seems to be little practical benefit of training a surrogate model.
As long as data-driven methods rely on data that is generated by classical methods, the latter can neither be ``superseded'' nor are they -- in many cases -- slower when considered in the proper context.

\begin{figure}[h]
    \centering
    \includegraphics{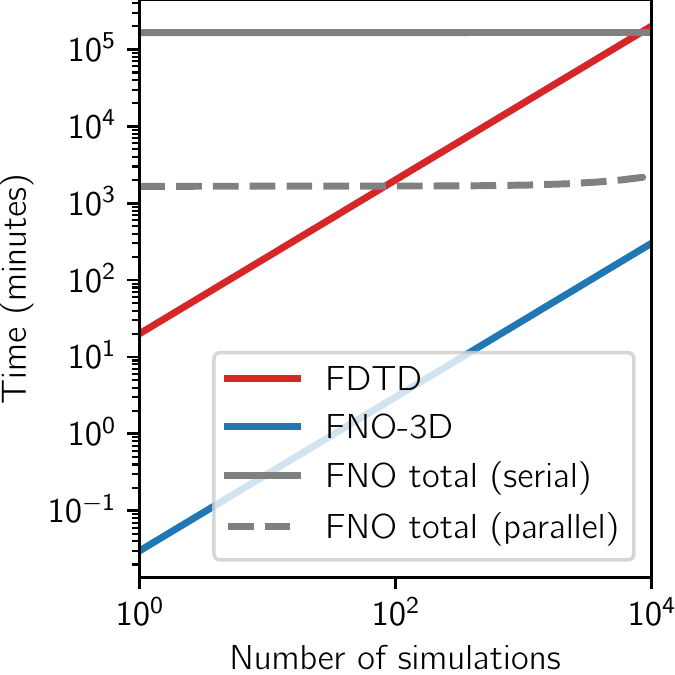}
    \caption{Comparison of the time required for full-wave simulations (FDTD) and FNO-3D predictions for a varying number of simulations. Time for a single simulation is assumed to be 20 minutes. Also shown is the total time of FNO-3D inference in addition to sample generation (8192 samples) and network training ($\approx 24$ hours) in the case of serial data generation (solid gray line) and parallel data generation (dashed gray line) used in this work.}\label{time}
\end{figure}

We illustrate this in~\cref{time}, where we show the total time taken for both FDTD and FNO-3D simulations and the time needed to generate FNO-3D training data.
Note that timings can vary considerably between different hardware and simulation setups, and the values shown in~\cref{time} should only be seen within the context of this work.
FNO-3D is indeed three orders of magnitude faster than FDTD during inference -- however, we argue that while this might be impressive when looked at in isolation, it is not a particularly relevant metric when making comparisons to conventional solvers.
The bulk of the time spent lies in data generation, in addition to training and inference time, both of which are practically negligible in comparison.
An advantage of data-driven approaches lies in the fact that individual samples are independent, \ie data generation can happen in parallel and is only practically limited by the computational resources at hand.
As an upper bound, and including training time, a surrogate model needs to be used for inference as many times as the number of samples contained in the dataset that it was trained on to just break even with a full-wave solver.
Beyond this point, a surrogate solver becomes cheaper concerning both total time and computational resources, \ie energy.
With respect to total time, this bound can be lowered considerably by generating samples in parallel as we have done here, see the~\nameref{data}.

All this is to say that data efficiency is \emph{crucial} in the context of scientific machine learning in general and for surrogate solvers in particular, and it needs to be addressed critically and transparently~\cite{woldseth2022use}.
We identify three aspects that should be considered when using deep learning in this context: \emph{architecture}, \emph{specialization}, and \emph{application}.
By choosing a suitable model architecture, the amount of training data required to reach sufficiently small test errors can be reduced significantly, as we demonstrate in the~\nameref{forward} for an electromagnetic scattering problem, and similar observations have been made across a variety of physical domains~\cite{cai2021deepm,lu2022multifidelity}.
Physics-informed approaches that incorporate governing equations into the model's loss function~\cite{wang2021learning,goswami2022physics,lim2022maxwellnet,chen2022high} similarly contribute to higher data efficiency.
Secondly, it is practical to aim for model specialization, \ie a surrogate solver does not need to generalize to all possible problem configurations, as such generalization tends to lead to unjustifiable data requirements.
To keep data requirements minimal, it is expedient to limit the scope of the model's intended application.
Lastly, the application needs to benefit from the use of a surrogate solver in a way that justifies the cost of training.
We demonstrate this in the~\nameref{inverse} by using a surrogate solver to perform a free-form, three-dimensional inverse design of electromagnetic scatterers.
Here, we train the model on a dataset of 8192 samples and run a total of 128 independent optimizations with 300 iterations each.
A comparable gradient-based approach using the adjoint method would equate to running $128 \cdot 300 \cdot 2 = 76800$ full-wave simulations, far exceeding the number of simulations performed for generating the training data.
Lastly, we would like to stress that while surrogate models \emph{can} be useful in many cases and will undoubtedly play an important role in the future, fast ``classical'', \eg semi-analytical, methods should generally be preferred if the problem at hand allows for their application as they often offer comparable speed at much higher accuracy and have well-controlled error bounds.

In summary, we have demonstrated the use of a neural operator-based model for solving electromagnetic scattering problems and show that it outperforms current state-of-the-art by a significant margin.
As a data-driven method, the surrogate solver inherently suffers from a loss of accuracy and generality when compared to full-wave solvers.
Nonetheless, we demonstrate that this approach can be used for complex tasks such as the free-form, gradient-based inverse design of three-dimensional electromagnetic scatterers.
Machine learning-based surrogate solvers have the potential to be highly useful for applications in nanophotonics.
However, it is essential to find efficient model architectures and identify tasks well-suited for their application.
We believe that our work contributes to both of these aspects and look forward to future developments in this field.

\section{Methods}

\subsection[data generation section]{Data generation}\label{data}

\begin{figure}[h]
    \centering
    \includegraphics{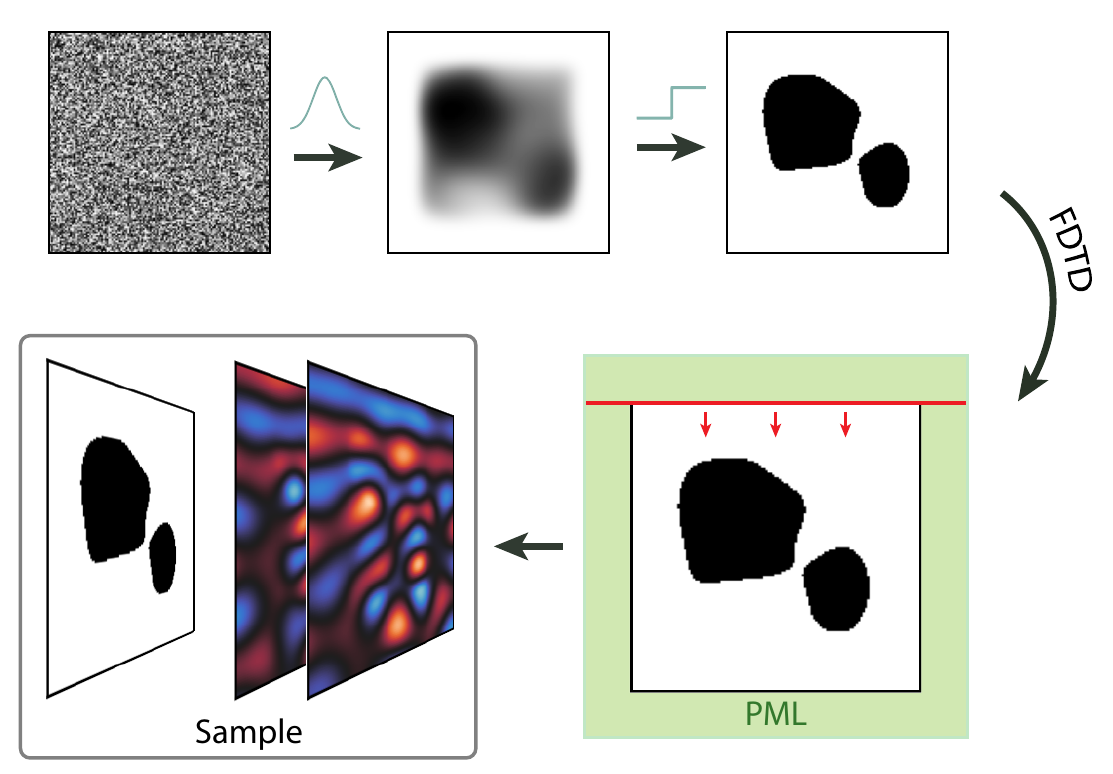}
    \caption{Illustration of the data generation process. A binary image is generated by smoothing and thresholding an image drawn from a random uniform distribution. The image is then interpreted as a material distribution and simulated with plane-wave illumination using FDTD. The sample for the training set is then assembled from the scatterer (input) and the individual field components (target).}\label{random_scatterers}
\end{figure}

To generate a large number of random scatterer geometries, we employ the same method as presented in~\textcite{repan2022exploiting}.
We sample points from a random uniform distribution on the interval $\interval[open right]{0}{1}$ on a regular square (cubic in 3D) grid with a side length of \SI{128}{\px}.
A zero-padded Gaussian blur ($\sigma = \SI{12}{\px}$) is then applied to the whole grid, and the result is thresholded at a value of $0.5$.
The zero-padding ensures that the scatterers are fully contained within the simulation domain and do not extend into the boundaries.
This procedure leads to a diverse set of smooth, random geometries containing one or multiple scatterers.
The whole data generation process is illustrated in~\cref{random_scatterers}.
We stress that the scatterers generated by this method can be almost arbitrarily complex, and the size of the Gaussian smoothing kernel primarily determines this complexity.
The size of the smoothing is directly analogous to the ``filtering'' method when choosing a minimal feature size in topology optimization~\cite{wang2011projection,sigmund2013topology}.

The generated random scatterers are interpreted as a material distribution, with $1$ indicating the presence of material ($n_{\text{high}} = 1.5$) and $0$ indicating air ($n_{\text{low}} = 1$), and illuminated using a plane-wave source at a wavelength of $\lambda_0 = \SI{1}{\um}$.
We choose $n_{\text{high}} = 1.5$ as it corresponds roughly to the typical polymers used in 3D laser nanoprinting.
We run the simulations at a spatial resolution of \SI{25}{\px\per\um}, which equates to a spatial extent of \SI{5.12}{\um} along each axis.
Additionally, the simulation domain is surrounded by perfectly matched layers (PMLs) with a thickness of \SI{0.5}{\um} on each side, increasing the side length of the simulation domain to \SI{6.12}{\um}.
The dataset comprises pairs of steady-state electric fields (excluding PML regions) and the corresponding scatterer.
The simulations for both the 2D and 3D datasets are performed using the open-source finite-difference time-domain (FDTD) software package Meep~\cite{oskooi2010meep}.
A summary of the datasets is given in~\cref{datasets}.

\begin{table}[h]
\centering
\caption{Summary of the datasets of random scattering geometries used for training all models.}\label{datasets}
\begin{tabular}{lrllll}
\toprule
Type & Samples & Input data & Output data & Input shape & Output shape \\
\midrule
2D & 17040 & Scatterer pixels & $E_z$ & $128 \times 128$ & $2 \times 128 \times 128$ \\
3D & 8720 & Scatterer voxels & $E_x$, $E_y$, $E_z$ & $128 \times 128 \times 128$ & $6 \times 128 \times 128 \times 128$ \\
\bottomrule
\end{tabular}
\end{table}

The total size of the 2D dataset is made up of $16384$ ($2^{14}$) training samples, $256$ validation samples, and $400$ test samples.
Smaller 2D datasets used in the~\nameref{forward} were sub-sampled from this larger dataset.

The 3D dataset is made up of $8192$ training samples, $128$ validation samples, and $400$ test samples.
Each sample took roughly 20 minutes (with some slight variations due to different scatterers) to simulate on four cores of an Intel Xeon Platinum 8368 CPU (76 cores total).
The data was generated on 40 nodes of the HoreKa cluster, with each node simulating 19 samples in parallel at a time.
In total, the generation of the 3D dataset took just under four hours.

\subsection[FNO \& UNet training section]{FNO \& UNet training}\label{fno_unet_training}

Details on the FNO hyperparameters are given in~\cref{fno_hparams}.
The UNet model consists of five downsampling (max pooling) and five upsampling blocks, where each block comprises six convolutional layers with batch normalization and ReLU activation.
We do not implement UNet ourselves in this work, instead, please refer to~\textcite{chen2022high} for a detailed overview of the architecture.
We deviate slightly from their model by replacing the periodic padding in the convolutional layers with zero padding, as we have found this to increase UNet performance on our dataset by $\approx \SI{1}{\percent}$ across all runs.

\begin{figure}[h]
    \centering
    \includegraphics{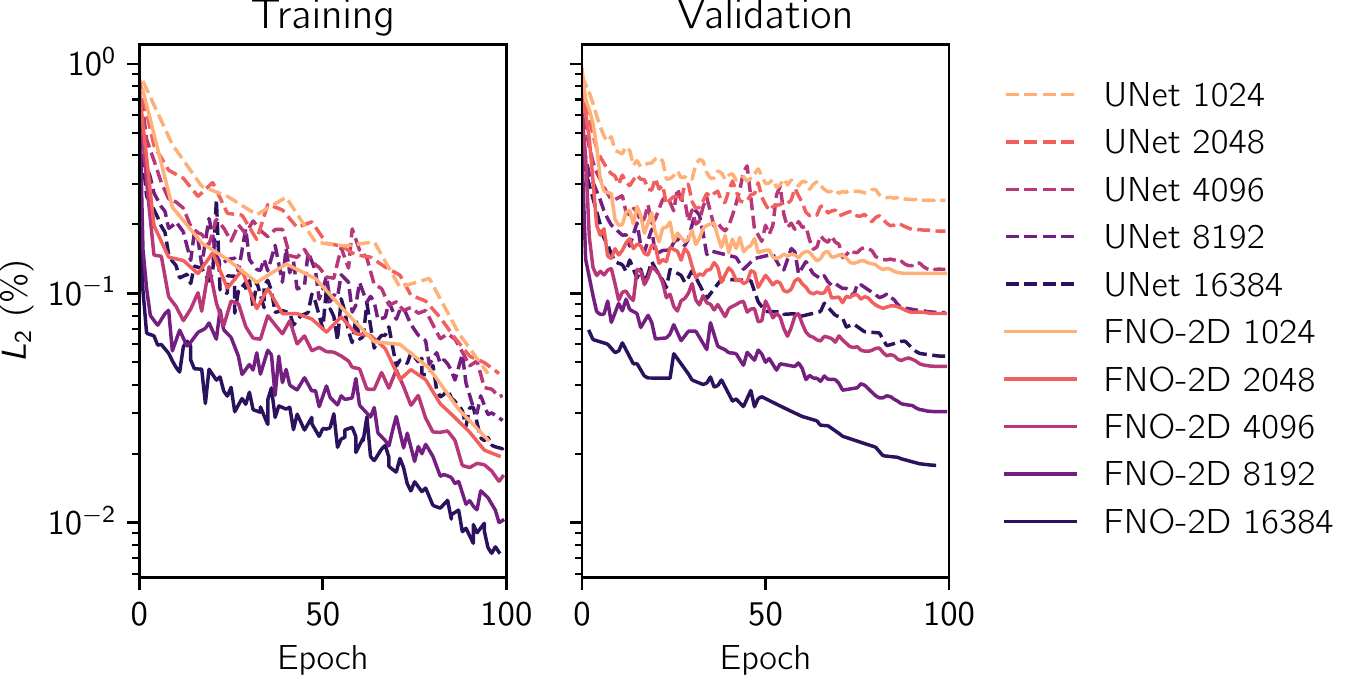}
    \caption{Training and validation $L_2$ loss curves for all UNet and FNO-2D training runs on different dataset sizes.}\label{training}
\end{figure}

All models are trained for 100 epochs using AdamW~\cite{loshchilov2017decoupled} with a one-cycle~\cite{smith2019super} learning rate policy and a batch size of 32.
We use the relative $L_2$ error (\cref{loss}, $p=2$) as the training loss function and monitor the relative $L_1$ and $L_2$ errors during validation at the end of each epoch with a validation split size of 256 samples.
Training and validation loss curves are shown in~\cref{training}.
The FNO-2D and UNet models are trained on a single NVIDIA A100 SXM4 GPU.
The training time per epoch is summarized in~\cref{nets_results} for each dataset size.

For training the FNO-3D model, we use the same effective batch size as for the 2D models.
This does not, however, fit into GPU memory during training, so we train the model in parallel across two nodes with 4 GPUs each, where each GPU only operates on 4 data samples.
To further reduce memory requirements, we also use activation checkpointing~\cite{fairscale2021} on the FNO blocks in the model.
After each iteration, the gradients are synchronized and averaged across all processes using PyTorch's \texttt{DistributedDataParallel} computing model~\cite{paszke2019pytorch}.
The total training time for the FNO-3D model was 11 hours.
The training hyperparameters are summarized in~\cref{train_params}.

\begin{table}[h]
\centering
\begin{threeparttable}
\caption{Training hyperparameters used for the surrogate solver models.}\label{train_params}
\begin{tabular}{lrrrr}
\toprule
Model & Parameters & Learning rate (min / max) & Batch size & Validation split \\
\midrule
UNet & \num{23617970} & \num{5e-5} / \num{5e-4} & 32 & 256\\
FNO-2D & \num{5909250} & \num{1e-3} / \num{1e-2} & 32 & 256 \\
FNO-3D & \num{141568902} & \num{1e-3} / \num{1e-2} & \si{2x4x4}\tnote{\textasteriskcentered} & 128 \\
\bottomrule
\end{tabular}
\begin{tablenotes}
\item[\textasteriskcentered] Trained on two nodes with 4 GPUs each and four samples per GPU (NVIDIA A100 SXM4).
\end{tablenotes}
\end{threeparttable}
\end{table}

Note that UNets are trained using a smaller learning rate than FNO -- this is because UNet training becomes unstable at a significantly lower learning rate than FNO.
To perform a fair comparison, we ran hyperparameter sweeps over multiple learning rates as well as learning rate schedulers for UNet.
The models presented herein always represent the best-performing networks that we found.

\subsection[VAE setup \& training section]{VAE setup \& training}\label{vae_training}

The VAE used in this work is based on a convolutional encoder-decoder type architecture.
The encoder contains five downsampling blocks that compress the input (a $128 \times 128 \times 128$ image of the material distribution) into a latent space representation.
Each downsampling block consists of a 3D convolution, a batch normalization layer, and SELU~\cite{klambauer2017self} activation.
For the convolutional layers, we use a kernel size of 5, a stride of 2, and a padding of 2 while doubling the number of channels in each layer.
The decoder mirrors the layers of the encoder, where the convolutional layers are replaced with transposed convolutions~\cite{zeiler2010deconvolutional} for upsampling.
The VAE network contains \num{86433514} trainable parameters in total.

The size of the latent space is, in principle, arbitrary and depends on the complexity of the geometries that should be modeled -- the more complex the geometries, the larger the latent space needs to be to represent all possible topologies.
We have found a latent space vector with 2048 elements to lead to good results during optimization for our random scattering geometries.
While VAEs with smaller latent sizes (down to around 256 elements) still show good reconstruction capabilities during training, we have found these to give worse results during inverse design, with less varied shapes and generally lower fidelity.
We suspect that while one might be able to reconstruct an input with a relatively small latent space adequately, the separation of different geometries within that space might be rather large and, thus, hard to reach for a gradient-based optimizer.
Suppose one chooses a larger latent space than what is strictly needed for reconstruction. In that case, the distance between different classes of geometries shrinks in this higher dimensional space, and an optimization might reach them more easily.

\begin{figure}[h]
    \centering
    \includegraphics{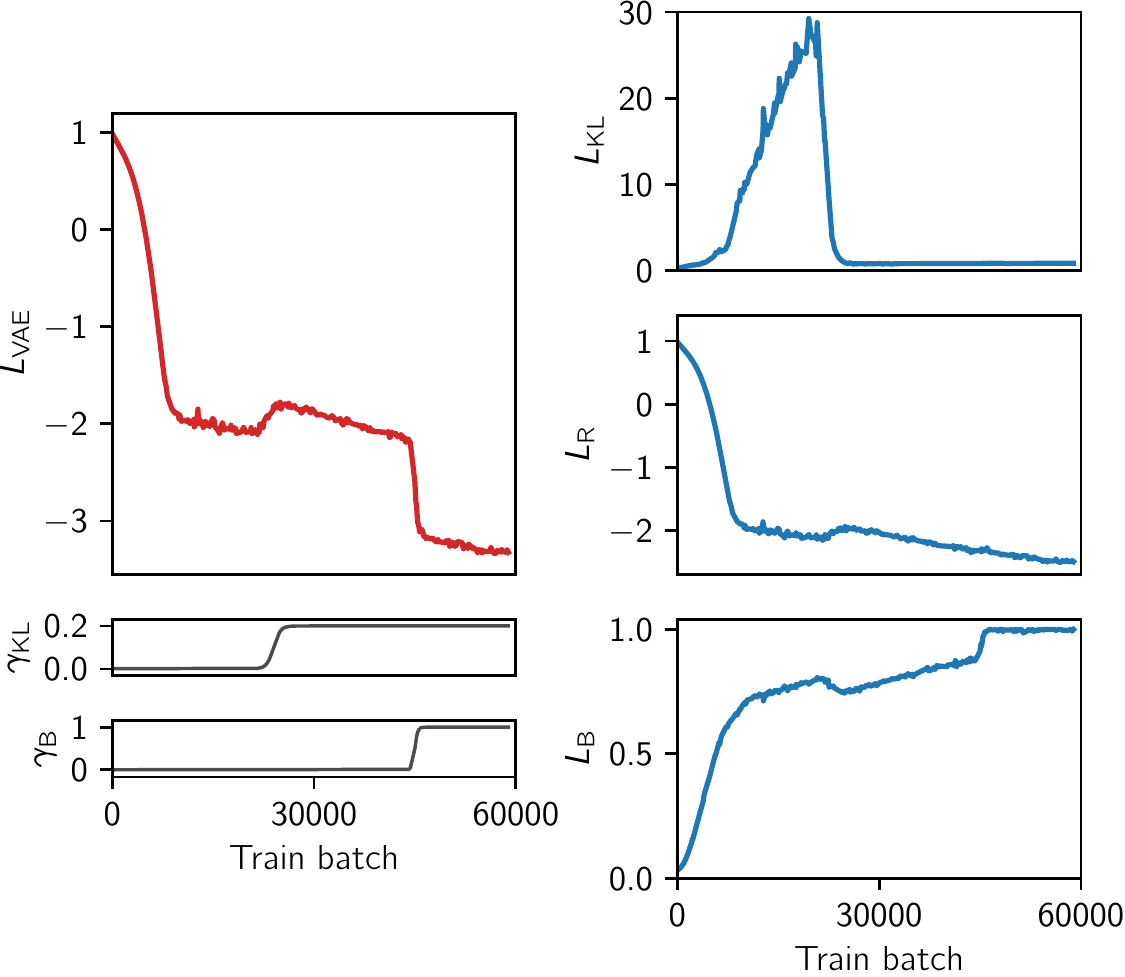}
    \caption{Total loss $L_{\text{VAE}}$ during training together with the annealed weight parameters $\gamma_{\text{KL}}$ and $\gamma_{\text{B}}$ (left) as well as the individual components of $L_{\text{VAE}}$ for KL-divergence $L_{\text{KL}}$, reconstruction $L_{\text{R}}$, and binarization $L_{\text{B}}$ (right).}\label{vae_loss}
\end{figure}

For training, we minimize the evidence lower bound (ELBO)~\cite{kingma2013auto} with an additional penalty term for binarization:
\begin{equation}
    L_{\text{VAE}}\qty(x, \hat{x}) = \gamma_{\text{KL}} \underbrace{L_{\text{KL}}\qty(x, \hat{x}) - L_{\text{R}}\qty(x, \hat{x})}_{\text{ELBO}} - \gamma_{\text{B}} L_{\text{B}} \qq{,}
\end{equation}
with the original image $x$, the reconstructed image $\hat{x}$, the Kullback-Leibler divergence $L_{\text{KL}}$, the reconstruction loss $L_{\text{R}}$ and the binarization penalty $L_{\text{B}}$:
\begin{alignat}{2}
    &L_{\text{KL}} &&= \mathbb{E}_q \qty[\log q \qty(z|x) - \log p \qty(z)] \\
    &L_{\text{R}} &&= \mathbb{E}_q \qty[\log p \qty(x|z)] \\
    &L_{\text{B}} &&= \frac{1}{3} \min\qty[-\log_{10}\qty(\frac{1}{N} \sum_i^N 4x_i \qty(1 - x_i)), \,3] \qq{,}
\end{alignat}
where $p$ is a standard normal distribution, $q$ is the distribution parametrized by the encoder output given a sample $x$, and $z$ is a sample drawn from $q$.
To balance the three terms of the VAE loss, we introduce the annealing parameters $\gamma_{\text{KL}}$ and $\gamma_{\text{B}}$ which increase the weighting of $L_{\text{KL}}$ and $L_{\text{B}}$ at different points in the training, respectively.
The training loss curves are presented in~\cref{vae_loss}.

We do not pre-generate a dataset for VAE training.
Instead, we use the geometry sampling procedure outlined in the~\nameref{data} to generate images of random scatterers on-the-fly during training.
The network is trained using a batch size of 64, for a total of \num{60000} batches, resulting in a total training time of 48 hours (single NVIDIA A100 SXM4 GPU).

\section[data availability section]{Data availability}\label{availability}

All code is made freely available under \url{https://github.com/tfp-photonics/neurop_invdes}.
This includes the code for data generation, network training, inverse design, as well as the model implementations.
We publish our research data under \doi{10.35097/911}, which includes the generated datasets and model weights as well as the code and data used for generating~\cref{nets_dists,fields,inverse_design_results,time,training,vae_loss}.


\section{Funding sources}

This research has been funded by the Deutsche Forschungsgemeinschaft (DFG, German Research Foundation) under Germany’s Excellence Strategy via the Excellence Cluster 3D Matter Made to Order (EXC-2082/1, Grant No. 390761711). Y.A. acknowledges support from the Carl Zeiss Foundation via the CZFFocus@HEiKA program.
T.R. is supported by the Estonian Research Council (Grant No. PSG716).
The authors acknowledge support by the state of Baden-Württemberg through bwHPC and the German Research Foundation (DFG) through grant no INST 40/575-1 FUGG (JUSTUS 2 cluster). The computations involved in this work were partially also performed on the HoreKa supercomputer funded by the Ministry of Science, Research and the Arts Baden-W\"urttemberg and by the Federal Ministry of Education and Research.

\section{Acknowledgements}

The authors kindly thank Lina Kuhn for helpful discussions regarding variational autoencoders.

\printbibliography

\end{document}